\documentclass[aps,prb,twocolumn,groupedaddress,showpacs]{revtex4}

\usepackage{graphicx}
\usepackage{dcolumn}
\usepackage{bm}
\usepackage{amsmath,amssymb}

\def\Vec#1{\mbox{\boldmath $#1$}}

\begin{document}


\title{Magnetization plateau of the $S=2$ antiferromagnetic Heisenberg chain with anisotropies}



\author{T\^oru Sakai,$^{1,2}$ Kiyomi Okamoto$^{1}$ and Takashi Tonegawa$^{3,4}$ }
\affiliation{$^1${Graduate School of Material Science, University of Hyogo, Kouto 3-2-1, Kamigori, Ako-gun, Hyogo 678-1297 Japan}\\
$^2${National Institute for Quantum and Radiological Science and Technology (QST) SPring-8, Kouto 1-1-1, Sayo, Sayo-gun, Hyogo 679-5148 Japan}\\
$^3${Professor Emeritus, Kobe University, Kobe 657-8501, Japan} \\
$^4$Department of Physical Science, School of Science, Osaka Prefecture University, Sakai 599-8531, Japan
}


\date{\today}

\begin{abstract}

We investigate the $S=2$ antiferromagnetic quantum spin chain with the 
exchange and single-ion anisotropies in a magnetic field, using the 
numerical exact diagonalization of finite-size clusters and the 
level spectroscopy analysis. 
It is found that a magnetization plateau possibly appears at a half of 
the saturation magnetization for some suitable anisotropy parameters. 
The level spectroscopy analysis indicates that the 1/2 magnetization 
plateau is formed by two different mechanisms, depending on the 
anisotropy parameters. 
The phase diagram of the 1/2 plateau states and some typical magnetization 
curves are also presented. 
In addition the biquadratic interaction is revealed to enhance 
the plateau induced by the Haldane mechanism. 

\end{abstract}


\maketitle


\section{INTRODUCTION}

Since Haldane predicted the spin excitation gap of the integer-spin 
antiferromagnetic Heisenberg chain,\cite{haldane1,haldane2}
the spin gap based on some topological 
nature has attracted a lot of interest. 
The existence of the Haldane gap was justified by many numerical studies.
\cite{botet,nightingale,sakai1,white,todo,nakano1,nakano2,wang}
Affleck, Kennedy, Lieb and Tasaki 
proposed a well-understandable picture of the spin gap formation, so-called 
the valence bond solid.\cite{aklt1,aklt2} 
The single-ion anisotropy $D$ tends to suppress the valence bond solid 
picture. When the anisotropy $D$ increases, a quantum phase transition 
occurs from the Haldane phase to the large-D phase where the topological 
nature disappears.\cite{sakai2,hida}
Recently Gu and Wen\cite{GuWen} and Pollmann et al.\cite{pollmann1,pollmann2}
introduced the concept of symmetry protected topological 
(SPT) phase to the quantum spin chain.
Based on their argument, the Haldane phase of $S=1$ chain is this SPT phase, 
while not in the case of $S=2$. 
On the other hand, the intermediate-$D$ phase even of $S=2$ chain 
predicted by Oshikawa\cite{oshikawa1} should correspond to the SPT phase. 
Unfortunately, 
early density matrix renormalization group calculation on 
the $S=2$ antiferromagnetic Heisenberg chain with the exchange 
anisotropy $\lambda$ and the single-ion one $D$ could not 
discover the intermediate-$D$ phase.
\cite{schollwoeck1,schollwoeck2,schollwoeck3} 
However, our recent study on the same $S=2$ model using the numerical 
exact diagonalization of finite-size clusters and the level 
spectroscopy analysis successfully detected the intermediate-$D$ phase.
\cite{tonegawa1, okamoto1, okamoto2, okamoto3, okamoto4} 
Since this phase appears only at a quite tiny region\cite{kjall,okamoto4} of the anisotropy 
parameter space, it would be difficult to discover it for some realistic 
materials. 
As another possibility to discover the SPT phase of the $S=2$ chain, 
we consider the magnetization process of the system. 
Since Oshikawa, Yamanaka and Affleck\cite{oya} discussed the magnetization 
plateau as the field induced Haldane gap, this problem has been 
investigated very extensively. 
Particularly the 1/3 magnetization plateau of $S=3/2$ chain 
was revealed to appear for sufficiently large $D$, by the numerical 
exact diagonalization study.\cite{sakai3} 
In addition the level spectroscopy analysis\cite{kitazawa} indicated that 
the intermediate-$D$ plateau phase, which corresponds to 
the SPT phase based on the VBS mechanism, as well as the large-$D$ 
plateau phase. 
Similar phenomena are expected to occur at half the saturation 
magnetization of the $S=2$ chain. 
In this paper, we consider the 1/2 magnetization state of the $S=2$ 
antiferromagnetic Heisenberg chain with the exchange and single-ion 
anisotropies using the numerical exact diagonalization of finite-size 
clusters and the level spectroscopy analysis, 
to discover the SPT phase, which corresponds to the intermediate-$D$ phase.

\section{MODEL}

Now we examine the magnetization process 
of the $S=2$ antiferromagnetic Heisenberg chain with the 
exchange and single-ion anisotropies, denoted by $\lambda$ and $D$, 
respectively. 
The Hamiltonian is given by 
\begin{eqnarray}
\label{ham}
&{\cal H}&={\cal H}_0+{\cal H}_Z, \\
&{\cal H}_0& = \sum _{j=1}^L \left[ S_j^xS_{j+1}^x + S_j^yS_{j+1}^y 
 + \lambda S_j^zS_{j+1}^z \right] \nonumber \\
  &&+D\sum_{j=1}^L (S_j^z)^2, \\
&{\cal H}_Z& =-H\sum _{j=1}^L S_j^z.
\end{eqnarray}
The exchange interaction constant is set to be unity as the unit of energy.
For $L$-site systems, 
the lowest energy of ${\cal H}_0$ in the subspace where 
$\sum _j S_j^z=M$, is denoted as $E(L,M)$. 
The reduced magnetization $m$ is defined as $m=M/M_{\rm s}$, 
where $M_{\rm s}$ denotes the saturation of the magnetization, 
namely $M_{\rm s}=L S$ for the spin-$S$ system. 
$E(L,M)$ is calculated by the Lanczos algorithm under the 
periodic boundary condition ($ \Vec{S}_{L+1}=\Vec{S}_1$) 
and the twisted boundary condition 
($S^{x,y}_{L+1}=-S^{x.y}_1, S^z_{L+1}=S^z_1$), 
up to $L=12$. 
Both boundary conditions are necessary for the level spectroscopy 
analysis. 

\section{MAGNETIZATION PLATEAU}

\begin{figure}
\includegraphics[width=0.95\linewidth,angle=0]{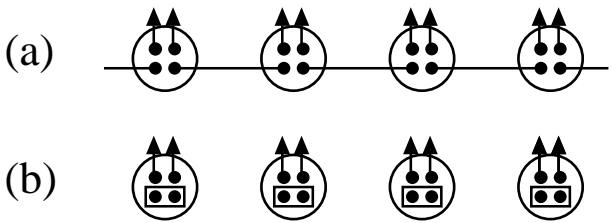}%
\caption{\label{vbs} 
Two different mechanisms of the 1/2 magnetization plateau; 
(a) Haldane mechanism and (b) large-$D$ mechanism. 
}
\end{figure}

Here we consider the state at $m=1/2$ in the magnetization process 
of the system (\ref{ham}) at $T=0$. 
In this state the magnetization per unit cell is $M/L$=1. 
Thus Oshikawa, Yamanaka and Affleck's theorem\cite{oya} suggests 
that the magnetization plateau possibly occurs
without the spontaneous breaking of the translational symmetry, 
because $S-M/L={\rm integer}$. 
If we consider the $S=2$ object as a composite spin consisting of 
four $S=1/2$'s, the 1/2 magnetization plateau is expected to 
appear due to two different mechanisms, as shown in Fig. \ref{vbs}. 
Namely one is (a) Haldane mechanism (a singlet dimer lies on each bond), 
and the other is (b) large-$D$ mechanism (the energy gap is open 
between the states $|S^z=1 \rangle$ and $|S^z=2 \rangle$ at each site 
due to the large $D$). 
The 1/2 magnetization plateaux based on the two mechanisms are 
called the Haldane plateau and the large-$D$ plateau, 
respectively in this paper. 
Following Pollmann et al.,\cite{pollmann1,pollmann2}
the SPT phase exists 
if any one of the following three global symmetries is satisfied: 
(i) the dihedral group of $\pi$ rotations about the $x$, $y$, and $z$ axes, 
(ii) the time-reversal symmetry $S_j^\mu \to -S_j^\mu$,
and (iii) the space inversion symmetry with respect to a bond.
It is easy to see that our Hamiltonian satisfies (iii),
but neither of (i) and (ii). 
Since the Tomonaga-Luttinger liquid phase is also possible, 
the $m=1/2$ state is expected to include the three phases; 
the Haldane plateau, the large-$D$ plateau and the gapless 
(plateauless) TLL phases. 

\section{LEVEL SPECTROSCOPY ANALYSIS}

\begin{figure}
\includegraphics[width=0.85\linewidth,angle=0]{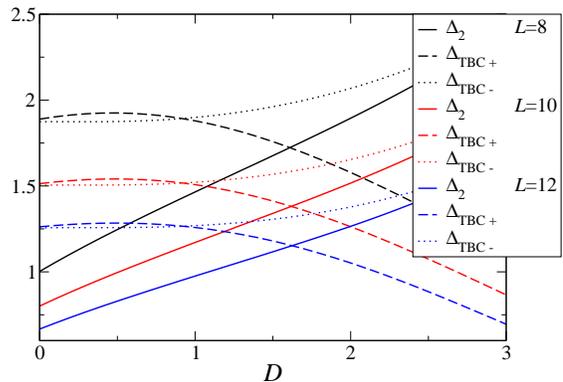}%
\caption{\label{LS10} 
Level spectroscopy analysis for $\lambda =1.0$. 
Solid, dashed and dotted lines are for $\Delta _2$, $\Delta _{\rm TBC +}$ 
and $\Delta_{\rm TBC -}$, respectively. 
Black, red and blue lines are for $L=$ 8, 10 and 12, respectively. 
}
\end{figure}

\begin{figure}
\includegraphics[width=0.85\linewidth,angle=0]{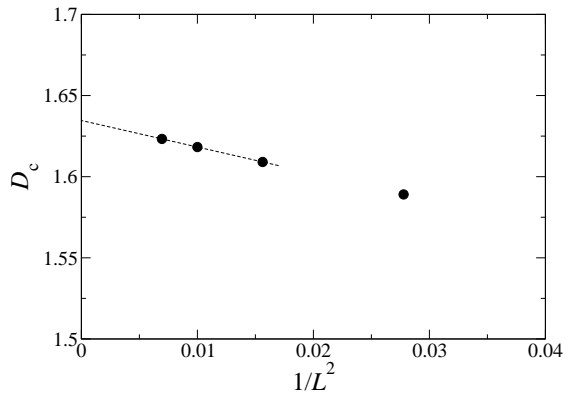}%
\caption{\label{extraD} 
Extrapolation of the critical value of $D$ between the TLL and large-$D$ plateau phases to the thermodynamic limit for $\lambda =1.0$. 
As the $L$ dependence of $D_c$ for the largest three system sizes 
is well fitted to $1/L^2$, $D_c$ in the thermodynamic limit 
is estimated by the least square method. 
}
\end{figure}

In order to distinguish these three phases, the level spectroscopy analysis
\cite{kitazawa} 
is one of the best methods. 
According to this analysis, 
we should compare the following three energy gaps; 
\begin{eqnarray}
\label{delta2}
&&\Delta _2 ={E(L,M-2)+E(L,M+2)-2E(L,M) \over 2}, \\
\label{tbc+}
&&\Delta_{\rm TBC+}=E_{\rm TBC +}(L,M)-E(L,M), \\
\label{tbc-}
&&\Delta_{\rm TBC-}=E_{\rm TBC -}(L,M)-E(L,M),
\end{eqnarray}
where $E_{\rm TBC+}(L,M)$ ($E_{\rm TBC-}(L,M)$) is the energy of the 
lowest state with the even parity (odd parity) with respect to the space inversion 
at the twisted bond under the twisted boundary condition, 
and other energies are under the periodic 
boundary condition. 
The level spectroscopy method indicates that the smallest gap 
among these three gaps for $M=L=M_{\rm s}/2$ determines the phase 
at $m=1/2$. 
$\Delta_2$, $\Delta_{\rm TBC+}$ and $\Delta _{\rm TBC-}$ 
correspond to the TLL, large-$D$-plateau and Haldane-plateau phases, 
respectively. 
The use of $\Delta_{{\rm TBC} \pm}$ directly reflects the above-mentioned (iii)
of the condition for the existence of the SPT phase.\cite{okamoto2}
The $D$ dependence of the three gaps calculated for $L=8$, 10 and 12 
is plotted for $\lambda =1.0$ in Fig. \ref{LS10}. 
It suggests that at the isotropic point ($\lambda$, $D$)=(1.0, 0.0) 
the system is in the TLL phase and increasing $D$ gives rise to 
a quantum phase transition to the large-$D$ plateau phase. 
The phase boundary is given by the cross point between $\Delta_2$ and 
$\Delta _{\rm TBC+}$. 
The system size dependence of the boundary is predicted to 
proportional to $1/L^2$, which is justified in Fig. \ref{extraD}. 
It indicates that the size correction of $D_{\rm c}$ is 
almost proportional to $1/L^2$, at least for $L=$8, 10, 12. 
Thus we estimate the phase boundary in the thermodynamic limit as 
$D_{\rm c}= 1.635 \pm 0.001$, fitting $1/L^2$ to the data for 
$L=$8, 10, 12. Unfortunately, the Haldane plateau phase does not 
appear for $\lambda=1.0$, different from $S=3/2$ chain.\cite{kitazawa} 

\begin{figure}
\vskip1cm
\includegraphics[width=0.85\linewidth,angle=0]{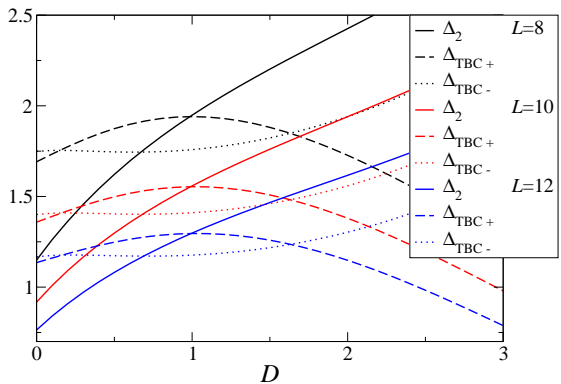}%
\caption{\label{LS20} 
Level spectroscopy analysis for $\lambda =2.0$. 
Solid, dashed and dotted lines are for $\Delta _2$, $\Delta _{\rm TBC +}$ 
and $\Delta_{\rm TBC -}$, respectively. 
Black, red and blue lines are for $L=$ 8, 10 and 12, respectively. 
}
\end{figure}

Next, the $D$ dependence of the three gaps is plotted for $\lambda =2.0$ 
in Fig. \ref{LS20}. In this case the Haldane-plateau phase appears 
between the TLL and large-$D$-plateau phases. 
The phase boundaries $D_{\rm c1}$ between TLL and Haldane phases and 
$D_{\rm c2}$ between Haldane and large-$D$ phases 
in the thermodynamic limit are estimated as 
$D_{\rm c1}=0.702 \pm 0.001$ and $D_{\rm c2}=1.633 \pm 0.001$, 
using the same fitting of $1/L^2$. 

\begin{figure}
\vskip1cm
\includegraphics[width=0.85\linewidth,angle=0]{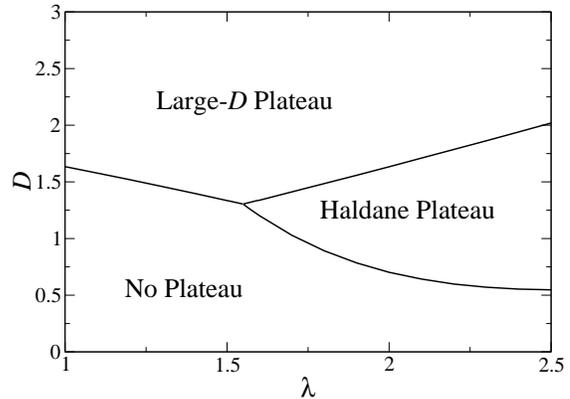}%
\caption{\label{phase} 
Phase diagram of the 1/2 magnetization state resulting from the 
level spectroscopy analysis. It includes the two plateau phases 
due to the Haldane and large-D mechanisms. 
}
\end{figure}

The phase diagram on the $\lambda$-$D$ plane is shown in Fig. \ref{phase}. 
It suggests that a tricritical point appears 
about ($\lambda$, $D$)=(1.55, 1.30). 
The Haldane-plateau phase would correspond to the SPT phase. 
Thus it should be called the symmetry protected topological plateau. 
This SPT phase appears in much wider region than that in the 
ground state phase diagram at $m=0$. 
Then the possibility of experimental discovery of the SPT phase 
for some real materials of the $S=2$ antiferromagnetic chain 
would be extended. 

\section{MAGNETIZATION CURVES}

\begin{figure}
\includegraphics[width=0.85\linewidth,angle=0]{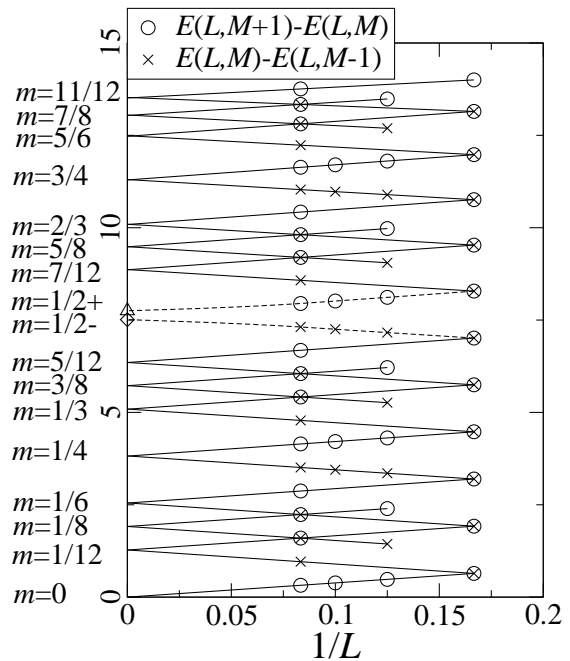}%
\caption{\label{extramag} 
$E(L,M+1)-E(L,M)$ and $E(L,M)-E(L,M-1)$ plotted versus $1/L$ with fixed $m$ 
for $\lambda=1.0$ and $D=2.0$. 
Each two quantities seem to coincide to the magnetic field $H$ for $m$ 
in the thermodynamic limit. The extrapolated points for $m=1/2+$ and 
$m=1/2-$ correspond to the result of the Shanks transformation 
$H_+(1/2)$=7.75 and $H_-(1/2)$=7.51, respectively. 
Dashed curves are guides for the eye. 
}
\end{figure}

Toward the experimental discovery of the 1/2 magnetization plateau, 
it would be useful to obtain the theoretical magnetization curve for some 
typical anisotropy parameters. 
In order to give the magnetization curve in the thermodynamic limit 
$L \rightarrow \infty$ using the numerical diagonalization results, 
we perform different extrapolation methods in the gapless and gapped cases. 
The magnetic fields $H_-(m)$ and $H_+(m)$ are defined as follows:
\begin{eqnarray}
\label{h1}
E(L,M)-E(L,M-1) \to H_-(m) \quad (L\to \infty), \\
\label{h2}
E(L,M+1)-E(L,M) \to H_+(m) \quad (L\to \infty), 
\end{eqnarray}
where the size $L$ is varied with fixed $m=M/M_{\rm s}$. 
If the system is gapless at $m$, the conformal field theory 
predicts that the size correction is proportional to $1/L$ and 
$H_-(m)$ coincides to $H_+(m)$.\cite{sakai4,sakai5} 
It is justified by Fig. \ref{extramag}, where $E(L,M)-E(L,M-1)$ and 
$E(L,M+1)-E(L,M)$ are plotted versus $1/L$ for $\lambda =1.0$ and 
$D=2.0$. 
It suggests that the system is gapless at $m \not=1/2$. 
The gapless feature at $m=0$ is consistent with the 
phase diagram of the previous work.\cite{tonegawa1} 
For these magnetization, we can estimate $H(m)$ in the thermodynamic 
limit, using the following extrapolation form
\begin{eqnarray}
\label{h0}
{1 \over 2}[E(L,M+1)-E(L,M-1)] \to H(m) + O(1/L^2).
\end{eqnarray}
On the other hand, if the system has a gap at $m$, namely 
the magnetization plateau is open, $H_-(m)$ does not 
coincides to $H_+(m)$ and $H_+(m)-H_-(m)$ corresponds to 
the plateau width. 
In such a case we assume the system is gapped at $m$ and 
use the Shanks transformation\cite{shanks,barber} 
to estimate $H_-(m)$ and $_+(m)$. 
The Shanks transformation applied for a sequence \{$P_L$\} is 
defined as the form
\begin{eqnarray}
\label{shanks}
P'_{L}={{P_{L-2}P_{L+2}-P_L^2}\over{P_{L-2}+P_{L+2}-2P_L}}.
\end{eqnarray}
As the above level spectroscopy analysis predicts that 
the 1/2 magnetization plateau appears for $\lambda=1.0$ and 
$D$=2.0, we use the method to estimate $H_-(m)$ and $H_+(m)$ 
at $m=1/2$. 
The Shanks transformation is applied for the sequence 
$E(L,M)-E(L,M-1)$ twice as shown in Table \ref{shanks1}.

\begin{table}[h]
   \caption{Result of the Shanks transformation applied for the sequence 
   $E(L,M)-E(L,M-1)$ twice. }
   \bigskip
   \begin{tabular}{|c|c|c|c|}
      \hline
      $L$& $P_L$ & $P_L'$ &$P_L''$ \\ \hline
      ~4~& ~6.7250103~ && \\ \hline
      ~6~& 7.0129442& ~7.3184395~ & \\ \hline
      ~8~& 7.1611715& 7.3918033& ~7.5105753~ \\ \hline
     ~10~~&  7.2514054 & 7.4371543  & \\ \hline
     ~12~~& 7.3121369 && \\ \hline
   \end{tabular}
   \label{shanks1}
\end{table}

\noindent
Within this analysis 
the best estimation of $H_-(1/2)$ in the thermodynamic limit 
is given by $P''_8$ and the error is determined by the difference 
from $P'_{10}$. 
Thus we conclude $H_-(1/2)=7.51 \pm 0.08$. 
The Shanks transformation applied for $H_+(1/2)$ is shown 
in Table \ref{shanks2}. 

\begin{table}[h]
   \caption{Result of the Shanks transformation applied for the sequence 
   $E(L,M+1)-E(L,M)$ twice. }
   \bigskip
   \begin{tabular}{|c|c|c|c|}
      \hline
      $L$& $P_L$ & $P_L'$ &$P_L''$ \\ \hline
      ~4~& ~8.6342191~ && \\ \hline
      ~6~& 8.2828303& ~7.9586157~ & \\ \hline
     ~8~& 8.1142027& 7.8716085& ~7.7518180~ \\ \hline
     ~10~~&  8.0147234 & 7.8212083  & \\ \hline
     ~12~~& 7.9490199 && \\ \hline
   \end{tabular}
   \label{shanks2}
\end{table}

\noindent
It gives the result $H_+(1/2)=7.75 \pm 0.07$. 
The estimated $H_-(1/2)$ and $H_+(1/2)$ for $\lambda=1.0$ and 
$D$=2.0 are shown as a diamond and a triangle, respectively in 
Fig. \ref{extramag} where dashed curves are guides for the eye.

\begin{figure}
\includegraphics[width=0.85\linewidth,angle=0]{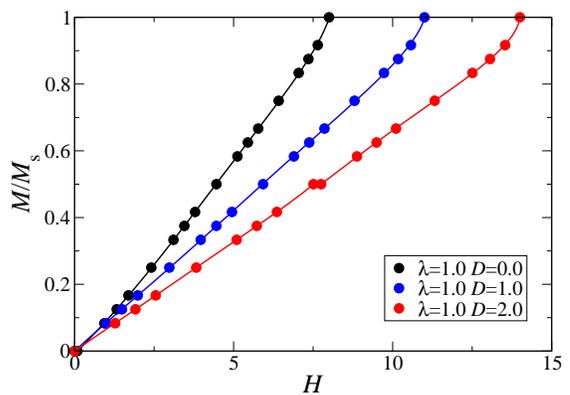}%
\caption{\label{mag10} 
Magnetization curves for $\lambda =1.0$ obtained by the numerical 
diagonalization and the extrapolation methods; the equation (\ref{h0}) 
for gapless points and the Shanks transformation for plateau points. 
The large-$D$ plateau appears at $m=1/2$ for $D=2.0$, 
while no plateau for $D=$0.0 and 1.0. 
Curves are guides for the eye. 
}
\end{figure}

\begin{figure}
\vskip1cm
\includegraphics[width=0.85\linewidth,angle=0]{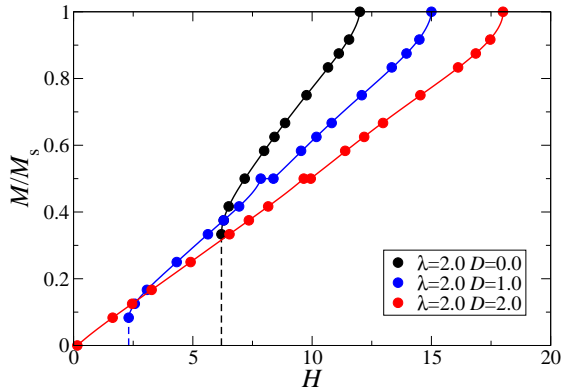}%
\caption{\label{mag20} 
Magnetization curves for $\lambda =2.0$ obtained by the same method 
as Fig. \ref{mag10}. 
The Haldane and large-$D$ plateaus appear for $D$=1.0 and 2.0, 
respectively. 
The magnetization jump from $m=0$ due to the spin flop transition also occurs 
for $D=0.0$ and 1.0. 
}
\end{figure}

Using these methods, the magnetization curves in the thermodynamic limit 
are presented for $\lambda =1.0$ ($D$= 0.0, 1.0 and 2.0) in Fig. \ref{mag10} 
and for $\lambda =2.0$ ($D$=0.0, 1.0 and 2.0) in Fig. \ref{mag20}.

In Fig. \ref{mag10} one of the precise estimations of the Haldane gap 
(0.0890)\cite{nakano2} is used as $H+(0)$ for $\lambda =1.0$ and $D=0$. 
As the ground state under $H=0$ for $\lambda =1.0$, $D=$1.0 and 2.0 
is in the $XY$ phase,\cite{tonegawa1} the magnetic excitation should be 
gapless. 
In Fig. \ref{mag20} the magnetization jump due to the spin flop 
transition occurs from $m=0$ for $D$=0.0 and 1.0, because the 
ground state under $H=0$ is in the N\'eel ordered phase.\cite{tonegawa1} 
As the precise magnetization curve around the jump is difficult to 
obtain by the numerical diagonalization, 
we assume that the magnetization jump occurs up to the smallest 
magnetization that is not skipped within the numerical diagonalization 
analysis. 
In any case the 1/2 magnetization plateau is quite small. 
Probably some precise magnetization measurement would be 
necessary to detect the 1/2 magnetization plateau of $S=2$ 
antiferromagnetic chain. 
If the Haldane plateau is too small to detect by the magnetization 
measurement, the ESR experiment to observe the edge spin effect 
at the doped impurity site\cite{hagiwara1} would be useful. 

\section{BIQUADRATIC INTERACTION}

\begin{figure}
\includegraphics[width=0.85\linewidth,angle=0]{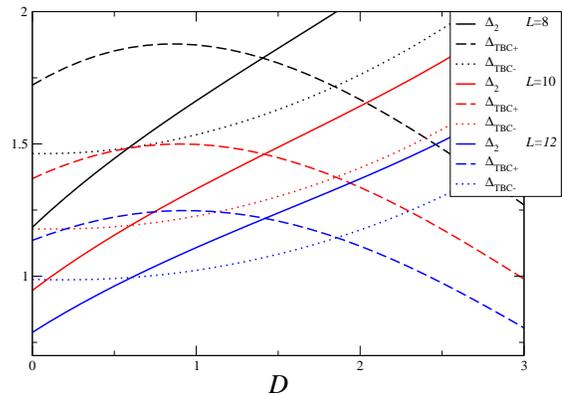}%
\caption{\label{LSBQ} 
Level spectroscopy analysis for $\lambda =1.0$ and $J_{\rm BQ}=0.05$. 
Solid, dashed and dotted lines are for $\Delta _2$, $\Delta _{\rm TBC +}$ 
and $\Delta_{\rm TBC -}$, respectively. 
Black, red and blue lines are for $L=$ 8, 10 and 12, respectively. 
It is found that the Haldane plateau phase appears even for $\lambda =1.0$. 
}
\end{figure}

\begin{figure}
\vskip2cm
\includegraphics[width=0.85\linewidth,angle=0]{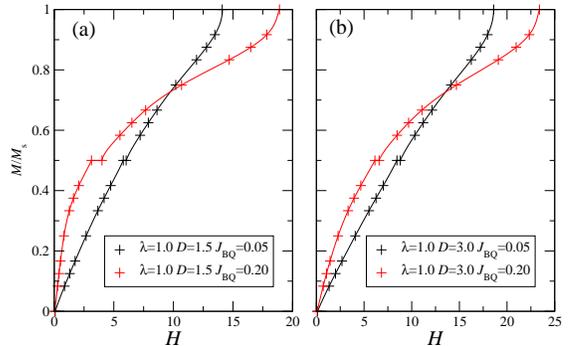}%
\caption{\label{magbq} 
Magnetization curves for $\lambda =1.0$ obtained by the same 
method as Figs. \ref{mag10} and \ref{mag20}: (a) for $D=1.5$ 
(Haldane plateau phase) and (b) for $D=3.0$ (large-$D$ plateau 
phase). Black and red curves are for $J_{\rm BQ}=0.05$ and 
$J_{\rm BQ}=0.20$, respectively. 
Curves are guides for the eye. 
When $J_{\rm BQ}$ increases, 
the Haldane plateau becomes much wider, while the large-$D$ 
plateau does not. 
}
\end{figure}

It would be important to consider the biquadratic interaction 
$J_{\rm BQ}\sum _j (\Vec{S}_j\cdot \Vec{S}_{j+1})^2$, 
because it possibly stabilizes the magnetization plateau.\cite{penc} 
The same level spectroscopy analysis as Figs. \ref{LS10} and \ref{LS20} 
is applied for the present model (\ref{ham}) including the 
biquadratic interaction. 
The result for $\lambda =1.0$ and $J_{\rm BQ}=0.05$ is shown in 
Fig. \ref{LSBQ}. 
It is found that the Haldane plateau phase appears even for $\lambda =1.0$, 
different from Fig. \ref{LS10}. 
The positive small biquadratic interaction is revealed to stabilize the Haldane plateau 
more than the large-$D$ one. 
Using the same method as Figs. \ref{mag10} and \ref{mag20}, 
the magnetization curves are given 
for $\lambda=1.0$ in Figs. \ref{magbq} (a) for 
$D=1.5$ (Haldane plateau phase) and (b) for $D=3.0$ (large-$D$ plateau phase), 
respectively. 
The magnetization curves for $J_{\rm BQ}=0.05$ and $J_{\rm BQ}=0.20$ are 
shown in Figs. \ref{magbq} (a) and (b), respectively. 
It indicates that the biquadratic interaction enhances the Haldane plateau, 
while not the large-$D$ one. 
Thus some materials including the biquadratic interaction 
would be better candidates to exhibit the Haldane plateau. 
Actually the level spectroscopy analysis indicates that 
the Haldane plateau appears for $J_{\rm BQ}>J_{\rm BQc}=0.0723$ 
even in the isotropic case ($\lambda =1$ and $D=0$). 
We hope the Haldane plateau will be discovered as the field induced SPT phase. 
One of the candidate materials of $S=2$ antiferromagnetic chain is 
MnCl$_3$(bpy).\cite{hagiwara2} However, the single-ion anisotropy $D$ was 
reported to be much smaller than the plateau phase of the present result 
and the biquadratic interaction is not expected to exist unfortunately. 

\section{SUMMARY}

In summary, the magnetization process of the $S=2$ antiferromagnetic 
Heisenberg chain with the exchange and single-ion anisotropies is 
investigated using the numerical exact diagonalization and the 
level spectroscopy analysis. 
As a result, the system possibly exhibits the 1/2 magnetization plateau 
due to Haldane mechanism, as well as the large-$D$ mechanism. 
The phase diagram of the $m=1/2$ state in the $\lambda$-$D$ plane is 
presented. The magnetization curves for several typical anisotropy 
parameters are also given. 
In addition the biquadratic interaction is revealed to enhance 
the Haldane plateau. 
We hope the present work would lead to the discovery of the 
field induced symmetry protected topological phase. 

\begin{acknowledgments}

This work was partly supported by JSPS KAKENHI, Grant Numbers 16K05419, 
16H01080 (J-Physics) and 18H04330 (J-Physics). 
A part of the computations was performed using 
facilities of the Supercomputer Center, 
Institute for Solid State Physics, University of Tokyo, 
and the Computer Room, Yukawa Institute for Theoretical Physics, 
Kyoto University.
\end{acknowledgments}


\end{document}